\documentstyle[12pt]{article}
\def\sp{\hspace*{.5 in}}
\def\beq{\begin{equation}}
\def\eeq{\end{equation}}

\def\dspace{\baselineskip = .30in}

\begin{document}

\title{Gauge Hierarchy, Planck Scale Corrections And The Origin of
         GUT Scale In Supersymmetric $(SU(3))^3$}

\author{{\bf G. Dvali}\thanks{Permanent address:  Institute of Physics,
Georgian Academy of
Sciences, 380077 Tbilisi, Georgia}\\
Dipartimento di Fisica, Universita di Pisa and INFN\\
Sezione di Pisa I-56100 Pisa, Italy\\
E-mail: dvali@mvxpi1.difi.unipi.it\\
{\bf Q. Shafi}\\
Bartol Research Institute, University of Delaware\\
Newark, DE 19716, USA\\
E-mail: shafi@bartol.udel.edu}

\maketitle
\date{ }

\begin{abstract} Within a supersymmetric unified framework we explore the
\linebreak resolution of the gauge hierarchy problem taking account of the
non-renormalizable terms in the superpotential.  For $[SU(3)]^3$ supplemented
by a discrete R parity, we find the remarkable property that the vacuum
configuration corresponding to the correct gauge symmetry breaking remains
flat (in the supersymmetric limit) to all orders in $M^{-1}_{Planck}$. The
grand unification scale arises from an interplay of the Planck and
supersymmetry breaking scales.  An `internal' $Z_3\otimes Z_4$ symmetry
protects a pair of electroweak doublets from becoming superheavy, yielding at
the same time the supersymmetric `$\mu$ term' with the right order of
magnitude. The $Z_4$ symmetry acts as matter parity and eliminates the
dangerous baryon number violating couplings. \end{abstract} \newpage

\dspace \section{Introduction}  In a recent letter, hereafter referred to as
[I], an attempt was undertaken to resolve the gauge hierarchy problem within a
specific grand unified theory, to wit, supersymmetric $G\equiv SU(3)_C \otimes
SU(3)_L \otimes SU(3)_R$ ($SU(3)^3$ for short). It was found that a
combination of discrete
symmetries suffices to ensure that a pair of electroweak higgs doublets
remains `light'.  More importantly perhaps, it was noted that a combination of
suitable discrete symmetries may provide an `all order' resolution of the
gauge hierarchy problem in G.  The main purpose of this paper is to address
this and the related problem of the origin of the GUT scale $M_G$ within the
framework of $G$.  It turns out that the $SU(3)^3$ approach does have some
unique features which we will highlight in this work.

        The resolution of the hierarchy problem in supersymmetric grand
unified theories (SUSY GUTS) would answer the following question:  How can a
theory with superheavy scales $M_G$ and $M_{Planck}$ ($=M_P$ for short) arrange
itself in such a way that a pair of light (mass $\sim M_W$) electroweak
doublets survive? There exist several mechanisms (missing partner [2],
pseudogolstone [3], custodial symmetry [4]), which (more or less naturally*
\footnote{*Criteria of ``naturalness'' vary in these approaches}) allow one to
obtain light doublets at the level of a renormalizable potential. However, in
SUSY GUTs (especially those based on superstrings or supergravity) there is no
obvious reason why one should restrict attention only to the renormalizable
couplings in the superpotential (and/or in the Kahler potential).  But then we
are immediately confronted by the problem of stability of the tree level
hierarchy in the presence of the higher order terms.  Even if kept `light'
(by some mechanism) at the tree level, the doublet mass can (and in general
will) be disturbed by the Planck scale operators (which, for the SUSY GUT
vacuum expectation values (vevs) cannot necessarily be regarded as small
corrections).

        Let us suppose that we want to solve the hierarchy problem in all
orders in a theory in which the only allowed input mass scales in the
superpotential are $M_P$ and $M_G$.  The scale $M_G$ either can be included in
the superpotential from the very beginning as an explicit input mass
parameter, or it can be dynamically or radiatively generated after SUSY
breaking.  In the latter case, in the unbroken SUSY limit, the scale $M_G$ is
represented by the undetermined (sliding) vev $<S>$ along the flat vacuum
direction, which gets fixed at $<S> = M_G $ only after SUSY breaking.

        A natural solution of the hierarchy problem in such a model would
imply that due to some symmetry, a pair of electroweak Higgs doublets $H
^{(1)}, H ^{(2)}$ has no mass term and/or renormalizable couplings with the
superlarge vev $<S>$ in the superpotential.  However, an absolutely decoupled
doublet can be problematic, since in the effective low energy theory we do
need a `small' $(\sim M_W)$ supersymmetric mass term $\mu H ^{(1)} H ^{(2)}$
for the doublets.  Thus, the doublets better couple to the vev $<S>$ at some
order in $M_P^{-1}$ .  Consider therefore the following term in the
superpotential: \beq {S ^{n+1} \over M_P^n} H^{(1)} H^{(2)}, n \geq 1 \eeq
After GUT symmetry breaking this operator will produce an effective $\mu$-term
of order $(M_G / M_P)^n M_G$.  With $M_G/M_P \sim 10 ^{-2}-10^{-3}$ we will
obtain the right order of magnitude for $n= 5 - 6$.

        Naively one might expect that this program is readily realized by
imposing some appropriate discrete or continuous symmetries (which of course
should be respected by the Planck scale physics).  However, there are two
major obstacles to be overcome. First of all, in ordinary GUTs (e.g. SU(5)
SO(10),....), the Higgs doublet is not an independent field but a member of
some irreducible representation of the GUT summetry.  This representation
usually includes a colored triplet component which can mediate rapid
\linebreak proton decay unless it is superheavy*.\footnote{*However, as was
shown in [5], this is not a necessary condition; in some cases the theory can
be arranged in such a way that the colored triplet is light but does not
mediate proton decay.}  Thus, we typically do need to couple this multiplet to
the large vevs, and the challenge is to keep the doublet(s) light.  This is
the famous doublet-triplet splitting problem in GUTs which often gets confused
with the hierarchy problem.

        It is difficult, however, to achieve this splitting to all orders,
since the symmetries which allow the renormalizable couplings needed for the
correct doublet-triplet splitting also permit nonrenormalizable terms with
rather low dimensionality that can give large masses to the doublets. Perhaps
the best motivated candidates for an `all order' resolution of the hierarchy
problem are theories in which the weak doublets and the colored triplets do
not reside in the same representation.  An explicit example of such a theory
is provided by $G\equiv SU(3)_C \otimes SU(3)_L \otimes SU(3)_R$.

        Another potential problem  is that the higher dimensional operators in
the superpotential cannot necessarily be regarded as small corrections since
they often provide a mechanism for the generation of the vevs, and can
strongly alter both the magnitude of the GUT scale as well as the symmetry
breaking pattern. The reason is that to suppress the dangerous low dimensional
operators as in (1), one often needs to assign an additional (discrete or
continuous) quantum number(s) to the superfield $S$ whose vev breaks the GUT
symmetry. However, this symmetry may also forbid the lowest (renormalizable)
selfcouplings of this superfield, allowing only the higher dimensional
invariants in the superpotential.

        To conclude, we find that the simplest candidate theories for an `all
order' solution of the hierarchy problem are the ones which satisfy the
following two conditions:

        (1)     Electroweak higgs doublets do not reside in the same
                representation as the colored triplets;

        (2)     All higher order invariants in the superpotential respect both
the correct pattern of the symmetry breaking, as well as the magnitude of the
GUT scale.  Indeed, we are particularly interested in models in which the GUT
scale appears after SUSY breaking, with the higher dimensional operators
playing an important role.

In addition,  we give preference to theories that do not require the
introduction of additional `auxilary' multiplets  which are not
otherwise necessary for the symmetry breaking. In
other words, the $SU(3)_C \otimes SU(3)_L \otimes SU(3)_R$ scheme
automatically fits in this class of desired theories.  The minimal Higgs
representation that can induce the correct symmetry breaking and also contain
the electroweak Higgs doublets are two pairs of $(1.\bar{3}.3) +
(1.3.\bar{3})$ representations.  Condition (1) above is automatically
satisfied since there is no colored component in this representation and thus
no doublet-triplet splitting problem.  As we have indicated in [1], by
supplementing $[SU(3)] ^3 $ with an appropriate discrete symmetry $Z_3 \otimes
Z_4 \otimes R$-parity, condition (2) can also be satisfied, thereby solving
the hierarchy problem in all orders!  More than that, it was argued that the
desired symmetry breaking pattern corresponds to a vacuum direction that is
unique and respected by all possible operators to all orders!  The SUSY GUT
scale arises from an interplay of the SUSY breaking parameters and higher
dimensional operators.

\section{Gauge Hierarchy to All Orders}

        Let us now discuss this question in more detail.  There are two pairs
of the Higgs superfields $\lambda ^A_{\alpha}, \bar{\lambda}^{\alpha}_A$ (and
$\lambda ^{\prime A} _{\alpha}, \bar{\lambda}^{\prime \alpha} _A$) transforming
as
$(1.\bar{3}.3) + (1.3.\bar{3})$ respectively.  Here and below we shall denote
by the Latin (A, B, C...= 1, 2, 3...) and Greek $(\alpha , \beta , \gamma = 1,
2, 3)$ symbols the $SU(3)_L$ and $SU(3)_R$ indices respectively.  For the
correct gauge summetry breaking $[SU(3)]^3 \rightarrow SU(3)_C \otimes SU(2)_L
\otimes U(1)$, the vevs should be oriented along the directions \beq \mid
\lambda \mid = \mid \bar{\lambda}^* \mid = \left[
\begin{array}{ccc}0&0&0\\0&0&0\\0&0&N\end{array}\right], \sp \mid \lambda
^{\prime} \mid = \mid \bar{\lambda} ^{\prime *} \mid = \left[
\begin{array}{ccc}0&0&0\\0&0&0\\0&\nu^{c \prime} &0\end{array}\right] \eeq Let
us find the minimal discrete symmetry which will guarantee that the vacuum
given by (2) is respected to all orders.  Surprisingly enough, it
turns out that all one needs is a discrete R-parity under which the chiral
superfields as well as the superpotential change sign.  This simply means that
the superpotential will only include odd power invariants in the superfields.
It is easily shown that ANY odd power invariant as well as its first
derivatives (with respect to the superfields) automatically vanish for the
configuration of the vevs given by (2).  Thus, all derivatives of the
superpotential are identically zero and the desired vacuum configuration is
respected to all orders!

        To prove this let us consider an arbitrary  $SU(3)_L \otimes SU(3)_R$
invariant product of the n-superfields transforming as $(1.\bar{3}.3)$ (such
as $\lambda, \lambda ^{\prime})$ and m-superfields transforming as
$(1.3.\bar{3})$ (such as $\bar{\lambda}, \bar{\lambda} ^{\prime})$.  Due to
R-parity the total number $(m+n)$ of superfields participating in this
invariant should be odd and thus $m \neq n$.  Any such invariant is obtained
from the direct product of the superfields \beq \underbrace{\lambda^{A_1}
_{\alpha _1} \otimes \cdots \lambda ^{A_n}_{\alpha_n}}_{n-times} \otimes
\underbrace{\bar{\lambda}^{\beta_1} _{B_1} \otimes \cdots \bar{\lambda} ^{\beta
_m} _{B_m}}_{m-times} \eeq with all possible $SU(3)_L \otimes
SU(3)_R$-invariant contractions of the indices. For simplicity we have dropped
the prime symbol in (3).  The number of primed and unprimed fields is not
important since they transform identically under the gauge group.  To form the
invariants, each lower $\alpha , B$ (upper $\beta , A$) index should be
contracted either with some other upper $\beta , A$ (lower $\alpha , B$ )
index or with the totally antisymmetric epsilon tensor.  But we know that $n
\neq m$ and the number of the upper and lower indices (for each $SU(3)_{L,R}$)
are different.  Thus, not all of them can be paired with each other.
Therefore, there should be at least one antisymmetric epsilon contraction
among the indices of each $SU(3)_{L,R}$.

        In other words, any appropriate invariant can be written in the
general form \beq \epsilon _{\alpha\beta\gamma}\ \epsilon _{ABC}\  f
_{\alpha\beta\gamma}, _{ABC} \eeq where $f _{\alpha\beta\gamma}, _{ABC}$ is
some odd product of the fields $\lambda , \bar{\lambda}, {\lambda} ^{\prime},
\bar{\lambda} ^{\prime}$ with all other indices contracted, and $\alpha ,
\beta , \gamma$ and $ABC$ are upper (or lower) indices belonging to the
components of $\lambda , \bar{\lambda} , {\lambda} ^{\prime} , \bar{\lambda}
^{\prime}$.  Of course, inside $f$ there can be some other epsilon
contractions too, but for us it is sufficient that there exists at least two
epsilon contractions which are explicitly indicated in (4).  Any such
invariant as well as its derivatives with respect to the fields $\lambda ,
\bar{\lambda}, {\lambda} ^{\prime}, \bar{\lambda} ^{\prime}$, will
automatically vanish along the configuration (2).  This is clear since along
this direction, the possible nonzero components are $\mid \lambda^3 _3 \mid =
\mid\bar{\lambda} ^3 _3 \mid \neq 0, \mid {\lambda} ^{\prime 2} _3\mid = \mid
\bar{\lambda} ^{\prime 3} _2 \mid \neq 0$. But, $\lambda ^3 _3 , {\lambda}
^{\prime 2} _3$ and $\bar{\lambda} ^3 _3 , \bar{\lambda} ^{\prime 3} _2$ share
one index, and therefore any epsilon contraction as well as its derivative
will automatically vanish along this direction.  Thus, the configuration (2)
corresponding to the right gauge symmetry breaking pattern automatically
ensures a vanishing first derivative of any existing invariant in the
superpotential, thereby satisfying to all orders and for any value of the
parameters the minimum (F-flatness condition):

\beq
{\partial W \over \partial \lambda} = {\partial W \over \partial \bar{\lambda}}
=
{\partial W \over \partial {\lambda} ^{\prime}} = {\partial W \over \partial
\bar{\lambda} ^{\prime}} = 0
\eeq

        Next let us ask the following question:  What is the minimal
`internal' discrete symmetry that can satisfy the following
requirements:

        (1)     Protect a pair of the electroweak doublets $H^{(1)}$, $H^{(2)}$
                from getting large masses;

        (2)     At the same time guarantee that there will appear
                a small supersymmetric mass term $\mu H^{(1)} H^{(2)}$ with
                $\mu \sim M_W$ .

        (3)     Allow large $(\sim M_G)$ masses for the other $SU(3)_C \otimes
SU(2)_L \otimes U(1)$-nonsinglet states, except, of course, the quark and
lepton superfields (and possibly some other states that may form  complete
SU(5) multiplets so as not to disturb unification of the gauge couplings).

        (4)     Eliminate the dangerous baryon number violating
                operators.

        It is remarkable that all of these requirements are met by an
appropriate $Z_4 \otimes Z_3$ symmetry.  Indeed we will now show that the $Z_4
\otimes Z_3$ symmetry is the unique `minimal' choice satisfying all of the
requirements (1) - (4) above. First, we need to identify those states that may
play the role of the electroweak Higgs doublets after $[SU(3)]^3$ breaks to
the standard model gauge group.  Inside the Higgs supermultiplets $\lambda ,
\bar{\lambda} , \lambda ^{\prime} , \bar{\lambda} ^{\prime}$ there are 12
electroweak doublet states.  Half of them ($\bar{H}^{(1)}, H^{(2)} , L ,
\bar{H}^{(1) \prime}, H^{(2) \prime}, L ^{\prime}$) carry quantum numbers of
the ``down''-type Higgs doublets, while the other half $H^{(1)} , \bar{H}
^{(2)} , \bar{L} , H ^{(1) \prime} , \bar{H}^{(2) \prime}, \bar{L} ^{\prime}$
carry quantum numbers of the ``up-''type Higgs doublet.  However, due to the
$[SU(3)]^3$ symmetry, only the ${H ^{(1)}} , {H ^{(2)}}$ pair (from $\lambda$
or $\lambda ^{\prime}$) has the correct Yukawa couplings with the quarks and
leptons.  The three generations of matter fermions (leptons, quarks,
antiquarks) transform as $(1.\bar{3} . 3), (3.3.1), (\bar{3}.1.\bar{3})$.  We
shall denote them as $\lambda _i, Q_i, Q^c_i$, where $i=1,2,3$ is the family
index.  It is obvious that the matter superfields do not have cubic
invariant couplings with $\bar{\lambda}$ or $\bar{\lambda} ^{\prime}$ fields
which transform as $(1.3.\bar{3})$.  The doublets $\bar{H}_{(1)},
\bar{H}_{(2)}, \bar{L}, \bar{H}_{(1)} ^{\prime} , \bar{H}_{(2)} ^{\prime} ,
\bar {L} ^{\prime}$ belonging to $\bar{\lambda} ^{\prime}, \bar{\lambda}$ do
not (at least at the renormalizable tree level) couple with the matter fields.

        The trilinear ``Yukawa'' couplings are of the form: \beq \lambda
\lambda _i \lambda _j  + \lambda Q_i Q^c_j  + \lambda ^{\prime} \lambda _i
\lambda _j  +  \lambda ^{\prime} Q_i Q^c_j \eeq Decomposing this into $SU(3)_3
\otimes SU(2)_L \otimes U(1)$-invariant pieces, we can easily find that
$H^{(2)} , H^{(2) \prime}$ and $H^{(1)} , H^{(1) \prime}$ doublets have the
correct down-type and up-type coupling with the quarks and leptons, just as in
the minimal SUSY standard model.  However, to minimize the number of `light'
doublets, we make the choice that only the pair $H^{(1)} , H^{(2)}$ be
protected by a discrete symmetry.  This is achieved by ensuring that the cubic
invariant $\lambda ^3$ (which contains the term ($H^{(1)}  H^{(2)}  N) $ and
is allowed by R-parity) is forbidden.  The minimal symmetry which accomplishes
this is $Z_2$, under which $\lambda \rightarrow - \lambda$.  Since the Yukawa
couplings of $\lambda$ with the matter superfields must be allowed, $\lambda
_i $ should transform under $Z_2$ in such a way that $(\lambda _i \lambda
_j)\rightarrow - (\lambda _i \lambda _j)$.  Here we assume that all the
families $(i,j = 1,2,3)$ transform in the same way under $Z_2$.  This
automatically fixes the transformation properties of the leptons, $\lambda _i
\rightarrow i \lambda _i (i=1,2,3)$.  It is natural to assume that the quarks
transform in the same way, namely, $(Q_i, Q^c_i) \rightarrow i (Q_i, Q^c_i)$.
Thus, it turns out that the symmetry which acts as a $Z_2$ on $\lambda$ , acts
as a $Z_4$ on the matter multiplets.  Note that this $Z_4$ symmetry also
forbids the dangerous baryon number violating operators which are trilinear in
the matter fields.  The proton is essentially stable in this theory.

        What are the transformation properties of the remaining Higgs
superfields $(\bar{\lambda} , \lambda ^{\prime} , \bar{\lambda} ^{\prime})$
under $Z_4$?  In order to prevent a proliferation of light doublets we should
allow the trilinear invariants \begin{equation} \bar{\lambda}^3, \lambda
^{\prime 3}, \bar{\lambda} ^{\prime 3} \end{equation} in the superpotential.
Thus $\bar{\lambda}, \lambda ^{\prime} , \bar{\lambda} ^{\prime}$ should not
transform under $Z_4$ . But this is not the end of the story.  The superfields
$\bar{\lambda}, \lambda ^{\prime} , \bar{\lambda} ^{\prime}$ must transform
under some other discrete symmetry, otherwise invariants of the form \beq
\lambda \lambda \lambda ^{\prime} +  {(\lambda \bar{\lambda}) \lambda ^3 \over
M^2_P}
+ ... \eeq will give very large masses to the doublets $H^{(1)}, H^{(2)}$ .
The only possible symmetry that allows trilinear invariants (7) is a $Z_3$
symmetry, and it is quite remarkable that this symmetry automatically forbids
(8)!  We will assume that the $\bar{\lambda}, \lambda ^{\prime} ,
\bar{\lambda} ^{\prime}$ fields transform under $Z_3$ as: \begin{equation}
(\bar{\lambda}, \bar{\lambda} ^{\prime}) \rightarrow  e^{i \alpha}
(\bar{\lambda} , \bar{\lambda} ^{\prime}) \sp and
\sp {\lambda}^{\prime} \rightarrow
e^{i2 \alpha} \lambda^{\prime}, \alpha = {2 \pi \over 3} \end{equation}

We now reach the important conclusion that the lowest dimensional coupling
which
gives rise to the `$\mu$-term' is given by

\beq {\lambda ^3 (\lambda \bar{\lambda})^3 \over M^6_P} \eeq which, for $\mid
\lambda \mid = \mid \bar{\lambda} \mid \sim M_G$, gives $\mu \sim \left(
{M_G \over M_P} \right) ^6 M_G $, as desired!

        In conclusion, starting from some fairly general considerations, we
are inevitably led to the unique discrete symmetry $Z_3 \times Z_4$ for
satisfying the four requirements listed below equation (5).

\section{Supersymmetry Breaking and the Generation of GUT Scale}

        We have shown in the previous section that in the supersymmetric limit
the configuration in (2) describes a valid vacuum in all orders (including all
possible renormalizable and nonrenormalizable terms), for arbitrary values of
the parameters compatible with R-parity.  The allowed invariants (as well as
their derivatives) identically vanish for this configuration of the vevs, so
that the magnitude of the vevs are not fixed, and our ``correct'' vacuum
corresponds to a flat direction.  The existence of ``accidental'' flat
directions in the vacuum is characteristic for supersymmetric theories.
However, the present case is a very unusual (and as far as we know the first
realistic) example of a flat direction that can survive to all orders in
$M^{-1}_P$ .  Consequently, the GUT scale in our theory can never be fixed (in
any order in $M^{-1}_P$) in the unbroken SUSY limit.  It is determined only
after supersymmetry breaking has occurred, offering as a consequence the
exciting possibility of explaining why $M_G$ happens to be close to $M_P \sim
10 ^{18} GeV$.  [There are previous examples (e.g. see [6], [7], [4]) in which
a sliding scale in the unbroken SUSY limit is fixed at $M_G$ after
supersymmetry breaking.]

        In conventional supergravity scenarios it is usually assumed that SUSY
breaking takes place in some ``hidden'' sector of the theory which
communicates with the ``observable'' fields ($\lambda , \bar{\lambda}, \lambda
^{\prime}, \bar{\lambda}^{\prime}, Q_i, Q^c_i, \lambda _i$) via some
nonrenormalizable Planck scale operators.  In the present work we do not wish
to advocate any particular mechanism which leads to SUSY breaking in the
hidden sector.  This breaking can be induced, for example, through the vev of
some fundamental scalar field (as in the simplest Polony potential [8]), or
dynamically due to some strongly coupled metacolor force (like in models with
gaugino condensation [9]).  For us, the important point about this breaking is
that it induces soft SUSY violating terms in the effective potential of the
observable fields, which in the minimal case has a well known form [10] \beq V
= \sum _{Z_{i}}\mid {\partial W \over \partial Z_i}\mid^2 + m^* _{3 \over 2}
A_n W ^{(n)} + h.c. + m^2 _{3 \over 2} \sum _{Z_{i}}\mid Z_i \mid ^2 +
[D-terms] \eeq Here $\mid m _{3 \over 2} \mid$ is a gravitino mass which
happens to be the messenger of SUSY breaking in the visible sector, and $A_n$
are $(n=1,2...)$ numbers, typically of order unity.  The $W^{(n)}$ denote the
n-linear pieces of the superpotential and the sum over $Z_i$ refers to all the
scalar components of the observable sector.

        As we have shown in section 1, all ${\partial W \over \partial Z_i}$
and all $W^{(n)}$ (for $n=1,...\infty)$, as well as D-terms, identically
vanish for the configuration of the vevs given by (2), corresponding to the
correct pattern of symmetry breaking.  Therefore, the only term that can
destabilize the vevs is $m^2 _{3 \over 2} [\mid N \mid ^2 + \mid \bar{N}
\mid^2 \linebreak + \mid \nu ^{c \prime} \mid ^2 + \mid \bar{\nu} ^{c \prime}
\mid ^2]$.  At the tree level (at scale $\sim M_P$), the quantity $m^2_{3
\over 2}$ is positive and thus the vevs are confined to the origin.  However,
it is well known that the sign of $m^2 _{3 \over 2}$ can change due to
renormalization effects, since ``Yukawa'' interactions (as well as cubic
couplings $\bar{\lambda}^3 , \lambda^{\prime 3} , \bar{\lambda}^{\prime 3}$)
can drive them negative.  If this happens, more precisely if $M^2_N + M^2
_{\bar{N}}< 0 , M^2_{\nu ^{c \prime }} + M^2 _{\bar{\nu}^{c \prime}} <0$ ,
then the potential (11) formally becomes unbounded from below and the minimum
is established at $<N>=\infty , <\nu ^{c \prime}>=\infty$. Fortunately this is
not the case, since for large enough vevs, one must take into account the back
reaction of the observable vevs on the hidden sector [the limit $M_P
\rightarrow \infty$ with $m_{3 \over 2} =$ fixed, under which the effective
potential (11) was obtained becomes invalid.]

        One possible mechanism which seems quite attractive assumes that the
spontaneous breaking of the R-parity (as well as SUSY) in the hidden sector
induces some R-parity noninvariant terms in the observable sector which will
help stabilize the GUT scale vevs. In general this will happen if the two
sectors communicate through the gravitational ($M^{-1}_P$ suppressed)
couplings in the superpotential (or at least in the Kahler potential). This
communication can be established, for instance,  via some additional gauge
singlet fields which play the role of `connectors': they pick up a nonzero
($R$-noninvariant) VEV from tree level interactions with the hidden sector
fields, and transfer this breaking to the observable sector through the Planck
scale couplings.

  Of course, it is very difficult to answer (without better knowledge of the
Planck scale physics) precisely what stabilizes the vevs at the high scales.
It is intriguing, however, to see how far we can go in understanding the
appearance of the GUT scale with our present knowledge of (super) gravity.  We
will see that under plausible assumptions, the vevs tend to be stabilized at
high scales which are somewhat below $M_P$.  A key point is that R-parity will
inevitably be broken in the hidden sector by the nonzero vev of the hidden
sector superpotential $<h>$ which changes sign under the R-parity
transformation.  Note that $<h>$ cannot vanish in the nonsupersymmetric
minimum with a zero cosmological constant.

  As we have mentioned before, the spontaneous breaking of the R-parity in the
hidden sector can (and in general will) induce effective R-parity noninvariant
terms in the observable sector through the general (super) gravity couplings.
On dimensional grounds it is clear that such terms are suppressed by powers of
$M_R / M_P$ (where $M_R$ is a $R$- noninvariant VEV communicating with the
observable sector).  Being noninvariant under R-parity, they sooner or later
lift the flat direction (2), and thereby stabilize the vevs at some high scale.

        To estimate the order of magnitude of these vevs, we need to know the
leading gravitational couplings between the hidden and observable sectors that
involve the R-parity breaking vevs.  Let us assume that the messenger of the
R-parity breaking is a gauge singlet $R$ which picks up the VEV through a tree
level interaction with the hidden sector fields.  This VEV, in general, can be
arbitrary. However, if we assume that these are the only two (input) scales
$M_P$ and $M_S \sim \sqrt {M_Pm_{3 \over 2}}$ in the hidden sector, then it is
natural to relate $<R>$ with one of them.  We will assume that $<R> \sim M_S$.

 The lowest dimensional couplings in the superpotential compatible with $Z_4
\otimes Z_3 \otimes$ R-parity are ${R(\lambda ^{\prime}
\bar{\lambda}^{\prime})^2 \over M^2_P}$ and ${R(\lambda \bar{\lambda})^6 \over
M_P^{10}}$ for the prime and nonprime sectors respectively. After the R-parity
breaking these will induce effective terms in the potential of the form: \beq
m_{3 \over 2}{\mid \lambda^{\prime}\mid^6 \over M^3_P}\sp and \sp m_{3 \over2}
{\mid \lambda \mid ^{22} \over M^{19}_P} \eeq which will stabilize the vevs at
scales $<\nu^{c \prime}> \sim M_P \left({m_{3 \over 2} \over M_P}\right)^{1
\over 4}$ and \linebreak $<N> \sim M_P \left({m _{3 \over 2}
\over M_P}\right)^{1 \over 20} $ respectively.  Admittedly,
the $<N>$ vev is somewhat higher (and $\nu^{c \prime}$ somewhat lower?) than
what we would prefer \linebreak $(\sim 10^{16} GeV)$, but the important
message here is that the origin of SUSY GUT scale possibly can be understood
within the framework of supersymmetric $(SU(3))^3$ models.

\section{Conclusion}

The gauge symmetry $G \equiv SU(3)_C \times SU(3)_L \times SU(3)_R$ provides
an attractive alternative to `standard' SUSY GUTs such as $SU(5)$ or
$SO(10)$.  In contrast to the latter (however see [11]),
$G$ it seems can arise from superstring
theories.  From a practical viewpoint, it allows for an elegant (`all order')
resolution of the hierarchy problem with a minimal set of higgs
supermultiplets.  Furthermore, the SUSY GUT scale arises from an interplay of
the Planck and SUSY breaking scales, with the higher dimensional operators
playing an essential role. [Note that if the GUT scale is put in by hand, a
single R-symmetry is enough to solve the hierarchy problem with $G \equiv
SU(3)_C \times SU(3)_L \times SU(3)_R$. See [1] for details.]  Both the proton
and the lightest supersymmetric particle are stable in the approach we have
described.
[6~

\section*{Acknowledgement}

One of us (Q.S.) acknowledges partial support by the Department of Energy
Grant No. DE-FG02-91ER40626.  We would like to thank R. Barbieri and K.S. Babu
for discussions.

\section*{References}
\begin{enumerate}
\item G. Dvali and Q. Shafi, Bartol Preprint No. BA-93-47, (1993), to appear
in Phys. Lett. B.

\item B. Grinstein, {\it Nucl. Phys.}, {\bf B206}, 387, (1982); \\ H.Georgi,
{\it Phys.Lett.}, {\bf B108}, 283, (1982);\\ A. Masiero, D. Nanopoulos, K.
Tamvakis and T. Yanagida, {\it Phys. Lett.}, {\bf B115}, 380, (1982);\\For
the ``missing vev'' variant of this mechanism in the context of SO(10) GUT see:
S. Dimopoulos and F. Wilczek, Santa Barbara Preprint VM-HE81-71, (1981),
unpublished;\\For a
discussion in the context of $(SU(3))^3$ see G. Lazarides and C.
Panagiotakopoulos, Univ. of Thessaloniki preprint UT-STPD-2-93 (1993);\\ For
more recent discussions in $SO(10)$ see K.S. Babu and S.M. Barr, Bartol
Preprint No. BA-94-04, (1994);\\
R. Mohapatra, Univ. of Maryland Preprint (1994).

\item K. Inoue, A. Kakuto and T. Takano, {\it Prof. Theor. Phys.}, {\bf 75}
664, (1986);\\ A.A. Anselm and A.A. Johansen, {\it Phys. Lett.},
{\bf B200}, 331, (1988);\\ Z. Berezhiani and G. Dvali, {\it Sov. Phys. Lebedev
Inst.}, {\bf Rep. 5}, 55, (1989);\\ R. Barbieri, G. Dvali and A. Strumia, {\it
Nucl. Phys.}, {\bf B391}, 487, (1993).

\item G. Dvali, {\it Phys. Lett. B} {\bf 324}, 59, (1994).

\item G. Dvali, {\it Phys. Lett.}, {\bf B287}, 101, (1992).

\item E. Witten, {\it Phys. Lett.}, {\bf B105}, 267, (1981).

\item A. Sen, {\it Phys. Rev.}, {\bf D31}, 900, (1985).

\item J. Polonyi, Budapest Preprint KFKI-1977-93 (1977);\\
E. Cremmer, S. Ferrara, L. Girardello, and A. van Proeyen, {\it Nucl. Phys.},
{\bf B212}, 413, (1983).

\item For a review see e.g. D. Amati et al., {\it Phys. Rep.}, {\bf 162}, 169,
(1988);\\
H.P. Nilles, {\it Int. J. Mod. Phys.}, {\bf A5}, 4199, (1990).

\item R. Barbieri, S. Ferrara and C.A. Savoy, {\it Phys. Lett.}, {\bf B119},
343, (1982);\\
A.H. Chamseddine, R. Arnowitt and P. Nath, {\it Phys. Rev. Lett.}, {\bf 49},
970 (1982).

\item R.Barbieri, G.Dvali and A.Strumia,  preprint No. IFUP-TH 20/94 and
 hep-ph/9404278, (1994).
\end{enumerate}

\end{document}